\documentclass[aps,pra,twocolumn,10pt]{revtex4-1}
\usepackage{graphicx}
\usepackage{amsmath}
\usepackage{amssymb}
\usepackage{mathrsfs}
\usepackage{amsfonts}
\usepackage{dsfont}
\usepackage{dcolumn}% Align table columns on decimal point
\usepackage{bm}% bold math
\usepackage{booktabs}
\usepackage{tabularx}
\usepackage{textcomp}
\usepackage{color}
\usepackage{xcolor}
\usepackage{float}
\usepackage{paralist}

\providecommand{\abs}[1]{\left\lvert#1\right\rvert}

\providecommand{\ave}[1]{\langle#1\rangle}

\providecommand{\mean}[1]{\langle#1\rangle} % si usa come: \mean{A} = <A>

\newcommand{\br}{\mathbf{r}}
\newcommand{\brt}{\bm{\rho}}
\newcommand{\bkt}{\bm{\kappa}}

\newcommand{\by}{\mathbf{y}}
\newcommand{\bz}{\mathbf{z}}
\newcommand{\bp}{\mathbf{p}}
\newcommand{\bq}{\mathbf{q}}
\newcommand{\bn}{\mathbf{n}}

\newcommand{\bk}{\mathbf{k}}

\newcommand{\vE}{\mathbf{E}}
\newcommand{\vB}{\mathbf{B}}

\newcommand{\vP}{\mathbf{P}}
\newcommand{\vS}{\mathbf{S}}

\newcommand{\vp}{\mathbf{p}}
\newcommand{\vq}{\mathbf{q}}

\newcommand{\vk}{\mathbf{k}}
\newcommand{\bnabla}{\bm{\nabla}}
\providecommand{\vv}[1]{\overset{\text{\small$\bm\leftrightarrow$}}{#1}}
\newcommand{\di}{\mathrm{d}}

\newcommand{\hx}{\hat{\mathbf{x}}}
\newcommand{\hy}{\hat{\mathbf{y}}}
\newcommand{\hz}{\hat{\mathbf{z}}}

\usepackage[all]{xy}

\begin{document}
%
%\title{Generating light with transverse spin}
\title{The ubiquitous photonic wheel}
\author{Andrea Aiello$^{1,2}$}
\email{andrea.aiello@mpl.mpg.de}
\author{Peter Banzer$^{1,2}$}
\affiliation{$^1$Max Planck Institute for the Science of Light, G$\ddot{u}$nther-Scharowsky-Strasse 1/Bau 24, 91058 Erlangen, Germany}
\affiliation{$^2$Institute for Optics, Information and Photonics, University of Erlangen-Nuremberg, Staudtstrasse 7/B2, 91058 Erlangen, Germany}
\date{\today}
\begin{abstract}
A circularly polarized electromagnetic plane wave carries an electric field that rotates clockwise or counterclockwise around the propagation direction of the wave. According to the handedness of this rotation, its \emph{longitudinal} spin angular momentum density  is either parallel or antiparallel to the propagation of light.
 However, there are also light waves that are not simply plane and carry an electric field that rotates around an axis perpendicular to the propagation direction, thus yielding \emph{transverse} spin angular momentum density.
Electric field configurations of this kind have been suggestively dubbed ``photonic wheels''.
 It has been recently shown that  photonic wheels  are commonplace in optics as they occur in electromagnetic  fields confined  by waveguides, in strongly focused beams, in plasmonic  and  evanescent waves.
 In this work we establish a general theory of electromagnetic waves  {propagating along a well defined direction, which carry} transverse spin angular momentum density. We show that depending on the shape of  {these waves, the} spin density may be either perpendicular to the \emph{mean} linear momentum (globally transverse spin) or to the linear momentum \emph{density} (locally transverse spin). We find that the latter case generically occurs only for non-diffracting beams, such as the Bessel beams.
Moreover, we introduce the concept of \emph{meridional} Stokes parameters to operationally quantify the transverse spin density.  To illustrate our theory, we apply it  to  the exemplary cases of Bessel beams and evanescent waves.
 These results open a new and accessible route to the understanding,  generation and manipulation of optical beams with transverse spin angular momentum density.
\end{abstract}

\maketitle

\section{Introduction}\label{Intro}

\emph{Photons are light},  namely they are massless and, consequently, do not possess a rest frame. Therefore, although they are quanta of vector fields of spin $1$, photons have only two independent spin projections along the direction of propagation (\emph{helicity}) \cite{Poynting1909,WeinbergI}. In the classical optics realm this means that at a fixed point in space, there are only two independent directions of oscillation (\emph{polarization}) for the electric field carried by monochromatic light \cite{BW}.
For plane and paraxial waves propagating along a given axis, which here and hereafter will be identified with the  $z$-axis of a Cartesian reference  frame, such oscillations are perpendicular to the direction of propagation $z$. If the two components of the electric field  have equal amplitude and are $\pi/2$ out of phase, the light is said to be circularly polarized. If the amplitudes are not equal, the light is elliptically polarized \cite{Jackson}.

 However, monochromatic electromagnetic waves confined by cavities and waveguides \cite{MIT,HausBook}, strongly focused beams \cite{RichardsWolf1959,PhysRevA.11.1365}, plasmonic fields  and evanescent waves \cite{HuardBook,Risset19937,Jozefowski:07,PhysRevA.88.033831,PhysRevA.89.033841,PhysRevA.90.023842}, exhibit  complex spatial structures and the two independent directions of oscillation of  the electric field  are
 not necessarily perpendicular to the propagation direction $z$. In these cases there is a longitudinal electric field component oscillating along  $z$ and the light can be elliptically polarized in a  meridional plane, namely in a plane containing the $z$-axis \cite{BW}. Therefore, the electric
field vector can spin around an axis perpendicular to $z$, as a ``photonic wheel'' \cite{JEOS:RP13032}, thus generating a transverse spin angular momentum (AM) density  \cite{PhysRevLett.103.100401,PhysRevA.81.053838,PhysRevA.85.061801}.
  Light beams with a longitudinal electric field component are key to many relevant applications in physics and chemistry, such as   particle acceleration \cite{PhysRevA.41.3727,PhysRevA.44.2656}, single-molecules manipulation and spectroscopy \cite{Macklin12041996,PhysRevLett.85.4482}, high resolution near-field optical microscopy \cite{NovotnyMicroscopy}, strong focussing of light \cite{Quabis20001,PhysRevLett.91.233901} and nanophotonics \cite{Novotny}.

  As recently reviewed by several authors \cite{AielloNatPhot,BliokhNatPhot,BliokhRev2015}, transverse spin AM density manifests in many different forms depending on the specific wavefield configuration \cite{BliokhEvanescent,doi:10.1021/nl5003526,PhysRevX.5.011039,PhysRevLett.114.063901}.
Here, we establish a general theory of monochromatic light waves carrying transverse spin AM density, which permits us to describe  different phenomena in a consistent fashion. We begin with by considering generic optical waves with longitudinal components of the electric field vector. From the study of the dynamical properties of these waves, such as energy, linear momentum and spin AM densities, we find that the last may be transverse in an either global or local sense; the first or second case occurring when the spin density vector is perpendicular to either the mean direction of the linear momentum of the wave or to the linear momentum density. In both cases, we quantify transverse spin AM density  by means of suitably defined meridional Stokes parameters \cite{Jackson}. An intriguing and previously unnoticed connection between the theory of complex functions (see, e.g., \cite{Lang}) and the transverse spin AM density of light is revealed by studying the \emph{maximal} transverse spin AM density occurring along the so-called c lines of the electric field, where the polarization of the electric field is purely circular \cite{Nye1987,Clines}.
 Finally, with the purpose of illustrating our findings,
we work out two detailed examples involving non-diffracting Bessel beams and evanescent waves.

\section{Transverse spin AM density}\label{Theory}

\subsection{Notation}\label{notation}

In order to set the notation and the units used throughout this paper, we now briefly review some notable results from the works of Nye \& Hajnal \cite{Nye1987}, Nye \cite{Nye1991} and Berry \& Dennis \cite{BerryC}.

The observable electric  field $\vE_\text{real}(\br,t)$ of a monochromatic electromagnetic wave of angular frequency $\omega$, can be written as the real part of the  time-harmonic complex  vector field $\bm{\mathcal{E}}(\br,t)=\vE(\br) \exp\left( - i  \omega t \right)$, where $\vE(\br) = \vp(\br) + i \vq(\br)$ and
\begin{align}\label{RealE}
\vE_\text{real}(\br,t) =  & \; \operatorname{Re} \bm{\mathcal{E}}(\br,t) \nonumber \\
=  & \; \mathbf{p}(\br) \cos (\omega t) + \mathbf{q}(\br) \sin (\omega t).
\end{align}
Here, $\bp = \vE_\text{real}(\br,0)$ and $\bq= \vE_\text{real}(\br,\pi/(2 \omega))$ are the conjugate radii of the so-called \emph{polarization ellipse} centered at $\br$,  swept out by  $\vE_\text{real}(\br,t)$ as time elapses. The  vector $\vp \times \vq$ normal to the polarization ellipse is parallel to the electric part of the cycle-averaged spin AM density  $\vS_E$, that is
\begin{align}\label{SE}
\vS_{E}(\br) = & \; \operatorname{Im} \bigl( \vE^* \times  \vE \bigr) \nonumber \\
= & \; 2 \bigl( \vp \times \vq \bigr),
\end{align}
where, from now on, a prefactor $\varepsilon_0/(4 \omega)$ in front of energy, linear and angular  momentum densities, will be omitted.

In a homogeneous plane wave with real-valued wave vector $\bk$, the propagation direction is naturally defined by $\bk$ itself. Such definition is \emph{global}, in the sense that the wave vector $\bk$ is independent of the position vector $\br$. However, for any wave that is not simply plane, one needs to specify a \emph{local}, namely dependent on $\br$, propagation direction. Let us denote with $\bk_E(\br)$ the local wave vector giving the propagation direction of the electric field $\vE(\br)$ at $\br$. It is defined as
\begin{align}\label{POE}
\bk_E(\br) = & \; \operatorname{Im} \left[ \vE^* \cdot \left( \bnabla \right) \vE \right] \nonumber \\
= & \; \vp \cdot \left( \bnabla \right) \vq - \vq \cdot \left( \bnabla \right) \vp \equiv \vp \cdot \bigl( \vv{\bnabla} \bigr) \vq,
\end{align}
where the suggestive  notation
\begin{align}\label{Notation}
\mathbf{A} \cdot \left( \bnabla \right)  \mathbf{B} = \sum_{j=1}^3 A_j \, \left(\hat{\mathbf{x}} \frac{\partial B_j}{\partial x} +\hat{\mathbf{y}} \frac{\partial B_j}{\partial y} + \hat{\mathbf{z}} \frac{\partial B_j}{\partial z}\right),
\end{align}
has been used \cite{OpticalCurrents}.
There is a simple relation between the local wave vector $\bk_E(\br)$ and the cycle-averaged  linear momentum of the electric field $\vP_E$. This can be written as the sum of an orbital part, denoted  $\vP_{OE}(\br)$, and a spin part, indicated with   $\vP_{SE}(\br)$, that is $\vP_E=\vP_{OE} + \vP_{SE}$.
It turns out that $\bk_E(\br)$ coincides (apart from the prefactor $\varepsilon_0/(4 \omega)$) with  $\vP_{OE}(\br)$, while the spin part  can be simply defined in terms of the spin AM density $\vS_E$ as:
\begin{align}\label{PSE}
\vP_{SE}(\br) =  \frac{1}{2} \bnabla \times \vS_E = \bnabla \times \bigl( \vp \times \vq \bigr).
\end{align}

Another important dynamical property of the electric field $\vE(\br)$, is its cycle-averaged energy density that can be written in terms of $\bp,\bq $ as
\begin{align}\label{WEpq}
W_E(\br) = \omega \left( \bp^2 + \bq^2\right).
\end{align}
From this equation and Eq. \eqref{SE} it follows that, given a fixed value of $W_E(\br)$,  $\abs{\vS_{E}(\br)}$ is \emph{maximal} when $\vp \perp \vq $ and $\abs{\bp}= \abs{\bq}$. In this case  $\abs{\vS_{E}(\br)} =2 \,\bp^2$ and the wave is circularly polarized.

In many cases of practical importance as, for example, in laser beams, light propagates along a well defined global direction, here and hereafter identified with the $z$-axis of a Cartesian reference frame $\{ x,y,z \}$. In such a frame, the  \emph{mean} value of a dynamical quantity like energy or linear momentum,  is defined as the integral of the corresponding density over the cross-section of the beam. For example, the mean wave vector is given by
\begin{align}\label{meanKE}
\mean{\bk_E} = \int \bk_E(\br) \di x \di y ,
\end{align}
and, because of our choice of the reference frame, it is parallel to the $z$-axis.
A special situation occurs when $\bk_E(\br) = k_{E}(\br) \hat{\mathbf{z}}$. In this case the local and global propagation directions coincide, although the wave is not plane. We will see explicitly later in what circumstances  this happens.

In general, for optical waves that are not simply plane, the dynamical properties and polarization patterns of the electric and magnetic fields differ \cite{Nye1999,BerryC}. In the remainder, we will study explicitly the transverse spin AM density of the electric field $\vE$ only. However, all our conclusions keep their  validity  if we replace $\vE$ with the magnetic field $\vB$. This is possible because for monochromatic electromagnetic waves the electric and magnetic fields are ``uncoupled'', in the sense that the electric and magnetic parts of the electrodynamic Hamiltonian are separately time-independent.

\subsection{Particular solutions}\label{PS}

In vacuum, in absence of free charges and electric currents, the electric vector field  $\bm{\mathcal{E}}(\br,t) $ is purely transverse, namely $\bnabla \cdot \bm{\mathcal{E}}=0$, and satisfies the wave equation $\Box \bm{\mathcal{E}} = 0$. A  plane-wave solution of these equations takes the form $\bm{\mathcal{E}}(\br,t) =  \mathbf{n} (\bk) \exp(i \bk \cdot \br - i\omega t)$, where $\bk = k \hat{\bk}$ is the wave vector.
Here $k \geq 0$ is the wave number and  $\hat{\bk}$ denotes an either real- or complex-valued unit vector such that $\hat{\bk} \cdot \hat{\bk}=1$, where the ``dot'' symbol  between any two $3$D vectors $\mathbf{a}$ and $\mathbf{b}$, stands for $\mathbf{a} \cdot \mathbf{b} = a_x b_x + a_y b_y + a_z b_z$ (no complex-conjugation required).
 {When $\hat{\mathbf{k}}$ is complex-valued, as in the evanescent waves considered in sec. \ref{Evanescentfield}, it can be written as $\hat{\mathbf{k}} = \hat{\mathbf{k}}_R + i \hat{\mathbf{k}}_I$, where $\hat{\mathbf{k}}_R$ and $\hat{\mathbf{k}}_I$ denote the real and imaginary parts of the unit wave vector, respectively. In this case the orthogonality condition $\hat{\bk} \cdot \hat{\bk}=1$ entails $\hat{\bk}_R \cdot \hat{\bk}_I=0$ and $|\hat{\bk}_R|^2 - |\hat{\bk}_I|^2=1$ \cite{Jackson}}.
 Moreover, $\omega = c k$ is the angular frequency of the wave and $ \bn(\bk)$ is a  three-vector perpendicular to $\bk$, namely $\bk \cdot \bn(\bk) =0$.   In our preferred Cartesian reference frame $\{x,y,z\}$, the condition $\bk \cdot \bn(\bk) =0$ can be explicitly written as
\begin{align}\label{kdotn}
\bk \cdot \bn(\bk) = k_x n_x + k_y n_y + k_z n_z = 0.
\end{align}
Apart from the trivial solution $\bn(\bk) = (0,0,0)$,  Eq.  \eqref{kdotn} also admits  three elementary  solutions of the form $\bn_Z(\bk) = (k_y,-k_x,0)/k$, $\bn_Y(\bk) = (k_z,0,-k_x)/k$ and $\bn_X(\bk) = (0,k_z,-k_y)/k$, where one of the three Cartesian components of $\bn(\bk)$ is chosen to be zero. The first solution $\bn_Z(\bk)$ generates a plane-wave mode whose electric field is purely perpendicular to the $z$-axis \cite{Lekner01}.
The remaining two solutions $\bn_X(\bk), \bn_Y(\bk)$ generate plane waves with both longitudinal ($z$) and transverse (either $y$ or $x$, respectively) components of the electric field and are, therefore, interesting to us.
Consider, for example, the solution $\bn(\bk)= \bn_X(\bk) $ with $\bk \in \mathbb{R}^3$. In this case the electric vector field can be evidently written as
\begin{align}\label{vApw}
\bm{\mathcal{E}}(\br,t) = (1/k)(0,k_z,-k_y) \exp(i \bk \cdot \br - i \omega t).
\end{align}
As desired, $\bm{\mathcal{E}}(\br,t)$  has transverse and longitudinal components of amplitudes $ k_z/k$ and $- k_y/k$, respectively, but it does not possess yet the structure of a photonic wheel because
from Eq. \eqref{RealE} it follows that the observable electric field
\begin{align}\label{RealE3}
\vE_\text{real}(\br,t) = \bn_X(\bk)  \cos(\bk \cdot \br - \omega t)  ,
\end{align}
simply oscillates along the direction $\bn_X(\bk)$ and does not rotate. Moreover, $\vE_\text{real}(\br,t)$ propagates in the direction of  $\bk$ which, in general, is not parallel to the $z$-axis. In fact,
\begin{align}
\bk_E(\br) = \bk \left( 1 - \frac{k_x^2}{k^2}\right).
\end{align}
To remove this problem, we notice that propagation along the $z$-axis can be simply achieved by taking the sum of two plane waves \cite{PhysRevX.5.011039} that are \emph{mirror-symmetric} with respect to the $z$-axis, namely
\begin{align}\label{vApw3}
\bm{\mathcal{E}}^X(\br,t) = & \; \frac{1}{2} \left[\bn_X(\bk)\exp(i \bk \cdot \br )+\bn_X(\overline{\bk})\exp(i \overline{\bk} \cdot \br )\right] \nonumber \\
& \; \times \exp(-i \omega t)\nonumber \\
= & \; \left(  \zeta \, \hat{\mathbf{y}} - i \eta \, \hat{\mathbf{z}}\right)  \exp(i k_z z - i \omega t),
\end{align}
where  $\overline{\bk} = (-k_x,-k_y,k_z)$ and $\zeta = (k_z/k) \cos(x k_x + y k_y)$, $\eta = (k_y/k) \sin(x k_x + y k_y)$. As byproduct of this choice, we automatically get a $\pi/2$ phase difference between the transverse and the longitudinal components of $\bm{\mathcal{E}}(\br,t)$, which ensures the rotation of the field in the meridional plane $yz$. Moreover, the exponential factor $\exp(i k_z z - i \omega t)$ shows that $\bm{\mathcal{E}}(\br,t)$ effectively propagates along the $z$-axis. Therefore, both conditions required for the existence of a photonic wheel field configuration are satisfied.
 A straightforward calculation reveals that, as expected,
the spin AM density carried by  $\bm{\mathcal{E}}^X(\br,t)$ is transverse,
\begin{align}\label{SE2}
\vS_E^X(\br) = - 2 \, \eta \, \zeta \, \hat{\mathbf{x}} ,
\end{align}
and that the local wave vector is purely longitudinal,
\begin{align}\label{KX}
\bk_E^X(\br) ={ k_z \left( \eta^2 +   \zeta^2 \right)} \, \hat{\mathbf{z}},
\end{align}
as desired. A similar reasoning can be repeated for the field obtained by replacing ${\mathbf{n}}_X$ with ${\mathbf{n}}_Y$ in Eq. \eqref{vApw3}. In this case we obtain
\begin{align}\label{comp}
\vS_E^Y(\br) =  2 \, \xi \, \zeta \, \hat{\mathbf{y}}  , \quad \text{and} \quad \bk_E^Y(\br) ={ k_z \left( \xi^2 +   \zeta^2 \right)} \,\hat{\mathbf{z}},
\end{align}
where $\xi = (k_x/k) \sin(x k_x + y k_y)$.  It is worth noticing that $|\vS_E^X| = 0$ if $k_y=0$ (the two plane waves lay on the $xz$-plane) and $|\vS_E^Y| = 0$ if $k_x=0$ (the two plane waves lay on the $yz$-plane). This is  a direct consequence of the transverse nature of electromagnetic fields. However, \emph{arbitrary} transverse orientation of $\vS_E$ may be achieved by superimposing, for a given wave vector $\bk$, the fields $\bm{\mathcal{E}}^X(\br,t)$ and $\bm{\mathcal{E}}^Y(\br,t)$, to obtain
\begin{align}\label{vApwBoth}
\bm{\mathcal{E}}(\br,t) = \cos \theta \, \bm{\mathcal{E}}^X(\br,t)+ \sin \theta \, \bm{\mathcal{E}}^Y(\br,t),
\end{align}
with $\theta \in \mathbb{R}$. It is not hard to show that  in this case we have
\begin{equation}
\begin{split}\label{SandK}
\vS_E(\br)  =  & \; 2 \, \zeta \left(\xi \sin \theta + \eta \cos \theta \right) \left( - \cos \theta \, \hx +  \sin \theta \, \hy \right), \\
\vk_E(\br)  =  & \; k_z \bigl[\zeta^2 +  \left(\xi \sin \theta + \eta \cos \theta \right)^2 \bigr] \hz .
\end{split}
\end{equation}

\subsubsection{Meridional Stokes parameters}\label{MeridStokes}

From Eq. \eqref{SE2} it follows that the polarization ellipse centered at $\br$ and swept out by $\operatorname{Re}[ \bm{\mathcal{E}}^X(\br,t) ]$, is located on the  $yz$-plane. This means that we can fully characterize such an ellipse by introducing the position-dependent \emph{meridional Stokes parameters} $S_0(\br),S_1(\br),S_2(\br),S_3(\br)$, defined in terms of the vectors $\bp, \bq$, as \cite{Dennis02},
\begin{equation}
\begin{split}\label{StokesPQ}
S_0(\br)  =  & \; p_z^2 + q_z^2 + p_y^2 + q_y^2 , \\
S_1(\br)  =  & \; p_z^2 + q_z^2 - p_y^2 - q_y^2 ,  \\
S_2(\br)  =  & \; 2 \left( p_z p_y + q_z q_y \right) , \\
S_3(\br)  =  & \; 2 \left( p_z q_y - p_y q_z \right) .
\end{split}
\end{equation}
Specifically, for $\bm{\mathcal{E}}^X(\br,t)$ we find
\begin{equation}
\begin{split}\label{PQ}
\bp = & \; \hat{\mathbf{y}} \, \zeta  \cos(z k_z) + \hat{\mathbf{z}} \, \eta  \sin(z k_z), \\
\bq = & \;  \hat{\mathbf{y}}\,  \zeta  \sin(z k_z) - \hat{\mathbf{z}} \, \eta  \cos(z k_z).
\end{split}
\end{equation}
From these expression and Eqs. \eqref{StokesPQ}, we obtain after a straightforward calculation,
\begin{equation}
\begin{split}\label{AllStokesPQ}
S_0   =  & \; \eta^2  + \zeta^2  ,  \\
S_1   =  & \; \eta^2  - \zeta^2  ,   \\
S_2   =  & \; 0 , \\
S_3   =  & \; 2 \, \eta \, \zeta  .
\end{split}
\end{equation}
At points $\br$ where $\bp(\br) \cdot \bq(\br) = 0$ and $ \bp^2(\br) - \bq^2(\br) =0$, the wave is circularly polarized and the spin AM density is \emph{maximal}. This occurs when $\zeta = \pm \eta$, that is when $x k_x + y k_y = \pm \arctan(k_z/k_y)$, with $k_y \neq 0$, irrespective of the values of $z$. As we will see later, the form \eqref{AllStokesPQ} for the Stokes parameters is generic for propagation invariant light beams carrying transverse spin AM density. It should be noted that
the  expression in the last row of Eq. \eqref{StokesPQ} reveals, when compared with Eq. \eqref{SE2}, that $S_3(\br) = - 2(\bp \times \bq)_x = - [\vS_E^X(\br)]_x$.

In all elementary examples considered above, the spin AM density and the local wave vector are perpendicular everywhere, irrespective of the position $\br$,
\begin{align}
\bk_E(\br) \cdot \vS_E(\br)=0.
\end{align}
Therefore, we say that the spin AM density is transverse in a \emph{local} sense, which is a strong condition. Later, when we will study fields more general than $\bm{\mathcal{E}}^{X,Y}(\br,t)$,  we will find waves with  spin AM density   transverse in a \emph{global} sense, namely fulfilling the weaker  requirement
\begin{align}
\ave{\bk_E} \cdot \vS_E(\br)=0 = \hat{\mathbf{z}} \cdot \vS_E(\br),
\end{align}
where we have used the fact that according to our choice of the reference frame,  $\ave{\bk_E}$ is directed along $\hat{\mathbf{z}}$.

In addition to the few basic properties illustrated in this section, the two plane-wave fields $\bm{\mathcal{E}}^{X,Y}(\br,t)$ possess many other interesting characteristics,  thoroughly investigated in \cite{PhysRevX.5.011039}.
However, these are only particular solutions of Maxwell's equations exhibiting transverse spin AM density. Conversely, we are going now to consider  general electromagnetic waves carrying transverse spin AM density.

\subsection{General solutions}\label{GS}

Let us begin the study of perfectly general fields exhibiting photonic wheel structures, by noticing that Eq. \eqref{vApw3} can be rewritten as \cite{Stratton}
\begin{align}\label{vApw4}
\bm{\mathcal{E}}^X(\br,t) = \frac{i}{k}\left( 0, -\partial_z \phi, \partial_y \phi \right)\exp(-i \omega t),
\end{align}
where $\partial_z \phi = \partial \phi/\partial z$, etc., and we have defined
\begin{align}\label{psiK}
\phi(\br) =  & \; \frac{1}{2} \bigl[ \exp(i \bk \cdot \br )+ \exp(i \overline{\bk} \cdot \br )\bigr] \nonumber \\
 =  & \; \cos \left( \bkt \cdot \brt \right)  \exp(i z k_z),
\end{align}
with $\bkt = k_x\hx + k_y \hy$ and $\brt = x \hx + y\hy$  denoting the transverse parts of the wave and position vectors, respectively.

Instead of the particular two plane-wave field  $\phi(\br)$,  let us consider now an arbitrary  solution $\psi(\br)$  of the monochromatic Helmholtz equation,  $(\nabla^2 + k^2)\psi = 0$, and use it to build the more general electric vector field
\begin{align}\label{vApw5}
\bm{\mathcal{E}}^X(\br,t) = \frac{i}{k}\left( 0, -\partial_z \psi, \partial_y \psi \right)\exp(-i \omega t).
\end{align}
The $\bp,\bq$  vectors of this field can be  calculated from Eq. \eqref{RealE}. The result is
\begin{equation}
\begin{split}\label{PQ2}
\bp = & \; \frac{1}{k}\left( \hat{\mathbf{y}} \frac{\partial v}{\partial z} - \hat{\mathbf{z}} \, \frac{\partial  v}{\partial y}  \right), \\
\bq = & \;  \frac{1}{k}\left(- \hat{\mathbf{y}} \frac{\partial u}{\partial z} + \hat{\mathbf{z}} \, \frac{\partial  u}{\partial y}  \right),
\end{split}
\end{equation}
where $u(\br)$ and $v(\br)$  denote, respectively, the real and the imaginary part of $\psi(\br) = u(\br) + i \, v(\br)$. It is not hard to show from Eq. \eqref{SE} that the spin AM density is either automatically transverse and parallel to the $x$-axis or null, depending on the shape of the functions $u$ and $v$:
\begin{align}\label{SEbis}
\vS_E^X(\br) =  \frac{2}{k^2}\left(  \frac{\partial v}{\partial z} \frac{\partial u}{\partial y}- \frac{\partial u}{\partial z} \frac{\partial v}{\partial y} \right) \hx.
\end{align}
A lengthy but straightforward calculation shows that the local wave vector $\vk_E^X(\br)$ possesses nonzero $x$-, $y$- and $z$-components (we do not report their expressions here because they are cumbersome and not very illuminating) and, therefore, is not longitudinal. This means that, as anticipated, $\vS_E^X(\br)$ is not transverse in a local sense. However, it can be still transverse in a global sense providing that $\ave{\vk_E^X}$ is parallel to $\hz$. In order to determine what conditions the generic field $\psi(\br)$ must satisfy in order to yield $\ave{\vk_E^X}$  parallel to $\hz$, we first rewrite it in the standard \emph{homogeneous angular spectrum} representation,
\begin{align}\label{AngSpec}
\psi(\br) =   \frac{1}{2 \pi}\int \widetilde{\psi}(\bkt) \exp(i \bkt \cdot \brt) \exp(i z \,k_z) \di k_x \di k_y,
\end{align}
where $k_z = (k^2 - k_x^2- k_y^2)^{1/2}$, and $\widetilde{\psi}(\bkt)$ is the $2$D Fourier transform of $\psi(x,y,0)$, with $\widetilde{\psi}(\bkt)=0$ for $k^2 - k_x^2- k_y^2<0$ \cite{MandelBook}. Then, substituting Eq.  \eqref{AngSpec} into Eq.  \eqref{vApw5} and using Eqs. \eqref{POE} and \eqref{meanKE}, we obtain
\begin{align}\label{aveKE}
\ave{\vk_E^X} = &   \int \bigl|\widetilde{\psi}(\bkt)\bigr|^2\left( k^2 - k_x^2\right)  \nonumber \\
&  \times \left[ k_x \hx + k_y \hy +  \sqrt{k^2 - k_x^2- k_y^2} \,\hz\right] \di k_x \di k_y.
\end{align}
This equation shows that whenever the modulus square of the angular spectrum is mirror-symmetric about the $z$-axis, that is when $\bigl|\widetilde{\psi}(\bkt)\bigr|^2 = \bigl|\widetilde{\psi}(-\bkt)\bigr|^2$, the mean wave vector becomes parallel to the $z$-axis and equal to
\begin{align}\label{aveKE2}
\ave{\vk_E^X} =   \hz \,\int \bigl|\widetilde{\psi}(\bkt)\bigr|^2\left( k^2 - k_x^2\right)  \sqrt{k^2 - \bkt^2} \, \di k_x \di k_y,
\end{align}
where $\bkt^2 = k_x^2 + k_y^2$.
This is the main result of this section. We make three remarks.
\begin{enumerate}[(i)]
  \item The condition $\bigl|\widetilde{\psi}(\bkt)\bigr|^2 = \bigl|\widetilde{\psi}(-\bkt)\bigr|^2$ is just the (weaker) version of the mirror-symmetry property $\phi(\br)=\phi(\overline{\br})$ exhibited by the two plane-wave field \eqref{psiK}. Thus, the same geometric condition induces both local and global transverse spin AM density.
  \item From Eqs. \eqref{StokesPQ} and $\psi = u + i v$, it follows that
\begin{equation}
\begin{split}\label{StokesPQg}
S_0(\br)  =  & \; u_z^2 + v_z^2 + u_y^2 + v_y^2 , \\
S_1(\br)  =  & \; -u_z^2 - v_z^2 + u_y^2 + v_y^2 ,  \\
S_2(\br)  =  & \; -2 \left( u_z u_y + v_z v_y \right) , \\
S_3(\br)  =  & \; 2 \left( u_z v_y - u_y v_z \right) ,
\end{split}
\end{equation}
where $u_z = \partial u/\partial z$, $u_y = \partial u/\partial y$ etc. If $\psi(x,y,z)$ were an analytic (or, holomorphic) function of the complex variable $w = z + i y$, then its real and imaginary parts $u$ and $v$, respectively, would satisfy the so-called Cauchy-Riemann conditions \cite{Lang},
\begin{align}\label{CR1}
 \frac{\partial u}{\partial z} = \frac{\partial  v}{\partial y}, \qquad  \frac{\partial u}{\partial y} =
 -\frac{\partial  v}{\partial z} .
\end{align}
Substituting these expressions into  Eqs. \eqref{StokesPQg} we obtain, after a little of algebra,
\begin{align}\label{Stokes000}
S_1 = S_2= 0 \quad \text{and} \quad S_0 = S_3 = 2 \abs{\frac{\partial \psi}{\partial y}}^2 .
\end{align}
 $S_3/S_0=1$ means that the light is circularly polarized and the transverse spin AM density is maximal. A similar results for ordinary waves with \emph{longitudinal} spin AM density, was obtained by Leckner \cite{Lekner03}.

If $\psi(x,y,z)$ is not an analytic function (as it is usually the case), the  conditions  \eqref{CR1} can still be satisfied \emph{locally}, namely at some isolated points $\br_0, \br_1,$ etc.
In this case Eqs. \eqref{CR1} become the equations determining the transverse c lines of the electric  field \cite{Nye1987}, along which the light is circularly polarized in the $yz$-plane and the spin AM density achieves its maximum value. In these points, the tangent to the c lines is perpendicular to the $z$-axis. In other words,  transverse c lines lay on the $xy$-plane by definition.
  \item Just as we did for the elementary two plane-wave fields, also in the general case we can consider both $\bm{\mathcal{E}}^X$ and $\bm{\mathcal{E}}^Y$ fields and their linear combinations of the form
\begin{align}\label{vApw7}
\bm{\mathcal{E}}(\br,t) = & \; \cos \theta\frac{i}{k}\left( 0, -\partial_z \psi, \partial_y \psi \right)\exp(-i \omega t) \nonumber \\
& + \sin \theta\frac{i}{k}\left(  -\partial_z \psi, 0, \partial_x \psi \right)\exp(-i \omega t).
\end{align}
The derivation of the formulas for the mean wave vector and the transverse spin AM density is straightforward and, therefore, will not be reported here.
%
% FILES: Calculations_NJP_2_XY.nb and Calculations_NJP_1_symbolic.nb
%
\end{enumerate}

\section{Propagation invariant fields}\label{Inva}

In the previous section we considered both  particular  and  general scalar fields $\phi(\br)$ and $\psi(\br)$, respectively, as the building blocks of electric vector fields carrying transverse spin AM density. We have found that  the spin density of the electric field generated by  $\phi(\br)$ is transverse in a local sense, that is perpendicular to the linear momentum density (or, local wave vector) $\bk_E(\br)$. Conversely,  the spin density from $\psi(\br)$ is transverse in a global sense, namely  perpendicular to the mean linear momentum density  $\mean{\bk_E}$. However, the existence of locally transverse spin density is not restricted to elementary plane-wave fields only.
In fact, suitably prepared propagation invariant beams as Bessel and Mathieu beams,   (see, e.g., \cite{VanEnkandNienhuis,Bouchal,PhysRevA.71.033411,Koo2006} and references therein), also possess a locally transverse spin density.
To show this, consider the two monochromatic scalar fields
\begin{equation}
\begin{split}\label{ND10}
\psi^X(\br) = V(x,y) \exp(i k_z z), \\
\psi^Y(\br) = U(x,y) \exp(i k_z z),
\end{split}
\end{equation}
where $k \geq k_z \in \mathbb{R}$ and $U(x,y), V(x,y)\in \mathbb{R}$ are independent solutions of the transverse Helmholtz equation $(\partial_x^2 + \partial_y^2 +\kappa^2)F=0$, with $F=U,V$ and $\kappa^2 = k^2 - k_z^2$. These fields are dubbed propagation invariant (or, non-diffracting) because the ``intensity'' $|\psi^{X,Y}(\br)|^2$ does not depend on $z$. The electric vector field built from Eq. \eqref{vApw7} with the replacements $ \psi \cos \theta \to \psi^X$ and $ \psi \sin \theta \to \psi^Y$, is
\begin{align}\label{vApw8}
\bm{\mathcal{E}}(\br,t) = & \;\frac{i}{k}\left( 0, -\partial_z \psi^X, \partial_y \psi^X \right)\exp(-i \omega t) \nonumber \\
& + \frac{i}{k}\left(  -\partial_z \psi^Y, 0, \partial_x \psi^Y \right)\exp(-i \omega t) \nonumber \\
= & \left[ \frac{k_z}{k} \left(U \hx + V \hy \right) +\frac{i}{k}\left(  \frac{\partial U}{\partial x} +  \frac{\partial V}{\partial y} \right) \hz\right]\nonumber \\
& \times \exp(i k_z z -i \omega t).
\end{align}
This field propagates along the $z$-axis and has a longitudinal component out of phase with respect to the transverse ones; therefore, photonic wheel structures are possible. The spin AM density can be calculated from Eq. \eqref{SE} after determining $\bp,\bq$. The result is
\begin{align}\label{SE5}
\vS_E(\br)  =    \frac{2k_z}{k^2} \left(  \frac{\partial U}{\partial x} +  \frac{\partial V}{\partial y} \right) \left( V  \hx -  U \hy \right).
\end{align}
The local wave vector is evaluated with the help of Eq. \eqref{POE} and is equal to
\begin{align}\label{K2}
\vk_E(\br)  =  & \;  \frac{k_z}{k^2} \left[ k_z^2\left(U^2 + V^2 \right) + \left(  \frac{\partial U}{\partial x} +  \frac{\partial V}{\partial y} \right)^2 \right] \hz \nonumber \\
= & \; k_z \abs{\vE(\br)}^2 \hz,
\end{align}
with $\bm{\mathcal{E}}(\br,t) = \vE(\br) \exp(- i \omega t)$. As anticipated in the previous section, in this case $\vk_E(\br)$ is factorable in the product between a position-dependent scalar function and a constant vector, namely $\vk_E(\br) = k_E(\br) \hz$, with $k_E(\br) = k_z \abs{\vE(\br)}^2$.
Since the generic functions $U(x,y)$ and $V(x,y)$ depend on the transverse coordinates $x,y$ only, we conclude that the densities  $\vS_E(\br)$ and $\vk_E(\br)$ are orthogonal at all points on the $xy$-plane, irrespective of the propagation distance $z$.
\begin{figure*}[ht!]
\centerline{\includegraphics[scale=3,clip=false,width=1.7\columnwidth,trim = 0 0 0 0]{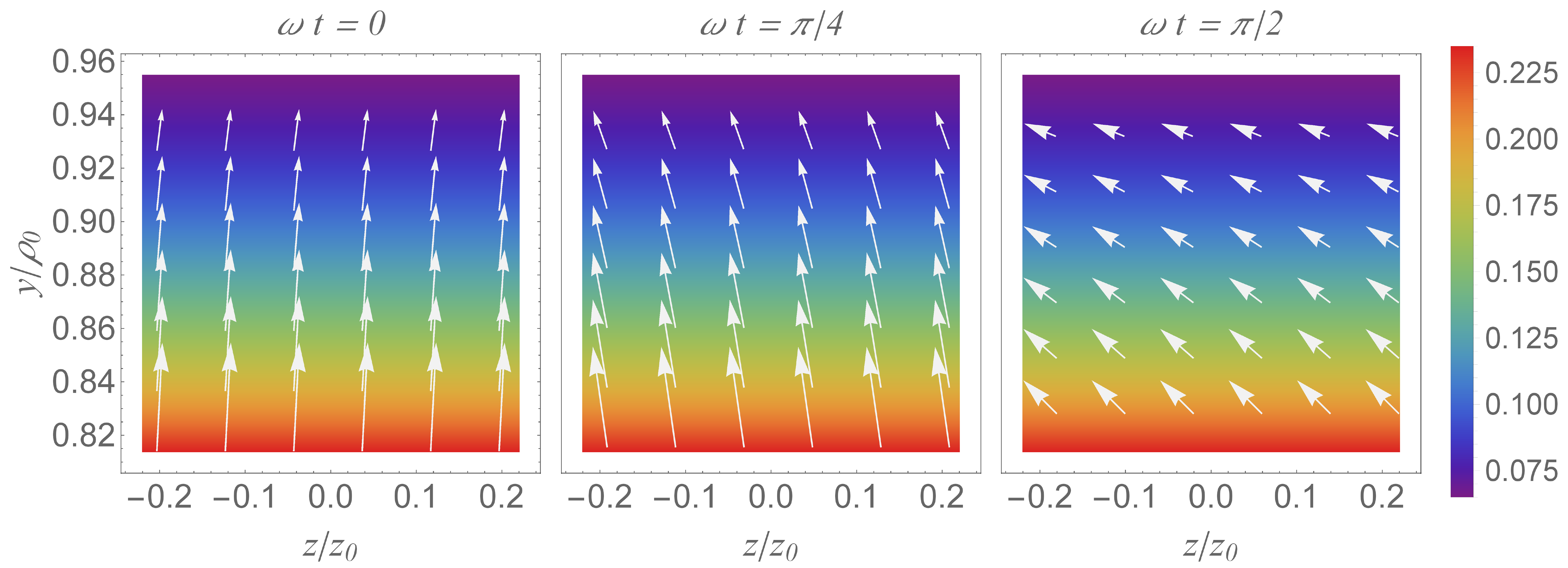}}
\caption{ \label{fig1}
Instantaneous distributions of the real-valued electric field $\vE_\text{real}(\br,t)$  \eqref{ERealBessel}  at $x=0$ and $y / \rho_0 \approx y_0$, where $y_0: \, \cot \vartheta_0   ={J_1\left(\kappa y_0 \right)}/{J_0\left(\kappa y_0 \right)}$, with $y_0 \simeq 0.884$ at $\vartheta_0 = 15^\circ$. At and around $(x,y)=(0,y_0)$ the electric field exhibits a circular motion in the $yz$-plane. Here  $\rho_0 = a/\kappa$ gives the central core spot size of the beam, with $a \approx 2.405$  such that $J_0(a)=0$ and $\vartheta_0 = 15^\circ$. Hues represent the magnitude $\abs{\vE_\text{real}(\br,t)}$ of the electric field normalized to $\abs{\vE_\text{real}(\bm{0},t)}$. From   \eqref{ERealBessel} it follows that beam appears linearly polarized along the $y$-axis when frontally observed on the $xy$-plane.
For comparison, see \cite{doi:10.1021/nl5003526,Grosjean:10}.
}
\end{figure*}

A convenient illustration of the spin and linear momentum densities \eqref{SE5} and \eqref{K2} is furnished by the fields
\begin{equation}
\begin{split}\label{Rad1}
U(x,y) = & \; \frac{x}{\rho} J_1(\kappa \rho), \\
V(x,y) = & \; \frac{y}{\rho} J_1(\kappa \rho),
\end{split}
\end{equation}
where $\rho = \sqrt{x^2 + y^2}$, $\kappa = \sqrt{k^2 - k_z^2}$, and $J_n$, with $n= 0,1,2, \ldots$, denotes a Bessel function of the first kind \cite{WolframJ}. The electric field \eqref{vApw8} is polarized in the meridional plane perpendicular to the azimuthal direction $\hat{\bm{\varphi}} = (-y\hx + x\hy)/\sqrt{x^2+y^2}$, that is
\begin{align}\label{vAp8}
\vE(\br) =   \left[\frac{k_z}{k} J_1(\kappa \rho) \hat{\bm{\rho}} + i \frac{\kappa}{k} J_0(\kappa \rho) \hz \right] \exp \left( i k_z z\right),
\end{align}
where $\hat{\bm{\rho}} = (x\hx + y\hy)/\sqrt{x^2+y^2}$.
 The spin AM density  is purely azimuthal and equal to
\begin{align}\label{SE6}
\vS_E(\brt)  =  -2 \left[\frac{\kappa}{k} J_0(\kappa \rho) \frac{k_z}{k} J_1(\kappa \rho) \right] \hat{\bm{\varphi}}.
\end{align}
 This expression has the same form of $\vS_E^X(\br) $ in Eq. \eqref{SE2} with $\eta \to {\kappa} J_0(\kappa \rho)/{k}$ and $\zeta \to {k_z} J_1(\kappa \rho)/{k}$. The main difference between the two densities is that $\vS_E^X(\br) $ has a Cartesian symmetry while $\vS_E(\bm{\rho})$ manifests cylindrical symmetry. The local wave vector is purely longitudinal and  given by  Eq. \eqref{K2}.

Electric fields polarized in the meridional plane perpendicular to the azimuthal direction, were considered in \cite{Clines} in the context of c lines in three dimensions. A paraxial version thereof was previously studied in \cite{Davis81}.

\section{Two simple examples}\label{Examples}

In this section we illustrate the theory developed above with the help of two clear cases. First, we consider a propagating zeroth-order Bessel beam with Cartesian symmetry. Then, we study an evanescent plane wave.

\subsection{Bessel field}\label{Besselfield}

Consider the scalar Bessel field
\begin{align}\label{Bessel}
\psi(\br) = J_0 \bigl( {k_r} \sqrt{x^2+y^2} \bigr) \exp\left(i z  k_z  \right),
\end{align}
where ${k_r} = k \sin \vartheta_0$ and $k_z = k \cos \vartheta_0$, with $\vartheta_0$ denoting the aperture of the Bessel cone \cite{PhysRevLett.58.1499}.
The electric field vector has the expression given in Eq. \eqref{vApw3} with
\begin{equation}
\begin{split}\label{Exa1}
\eta = & \; \frac{{k_r}}{k} J_1 \bigl({k_r} \sqrt{x^2+y^2} \bigr) \frac{y}{\sqrt{x^2+y^2}}, \\
\zeta = & \; \frac{k_z}{k} J_0 \bigl({k_r} \sqrt{x^2+y^2} \bigr).
\end{split}
\end{equation}

Transverse spin AM density and local wave vector can be calculated substituting  Eqs. \eqref{Exa1} into Eq. \eqref{SE2} and Eq. \eqref{KX}, respectively. They are directed along the $x$- and $z$-axis, respectively. The observable electric vector field is given by
\begin{align}\label{ERealBessel}
\vE_\text{real}(\br,t) =  & \;  \hat{\by} \, \frac{k_z}{k} J_0 \bigl({k_r} \sqrt{x^2+y^2} \bigr) \cos \left( z k_z - \omega t \right) \nonumber \\
&  + \hat{\bz} \, \frac{{k_r}}{k} J_1 \bigl({k_r} \sqrt{x^2+y^2} \bigr) \frac{y}{\sqrt{x^2+y^2}}\nonumber \\
& \times \sin \left( z k_z - \omega t \right) .
\end{align}
\begin{figure}[ht!]
\centerline{\includegraphics[scale=3,clip=false,width=.8\columnwidth,trim = 0 0 0 0]{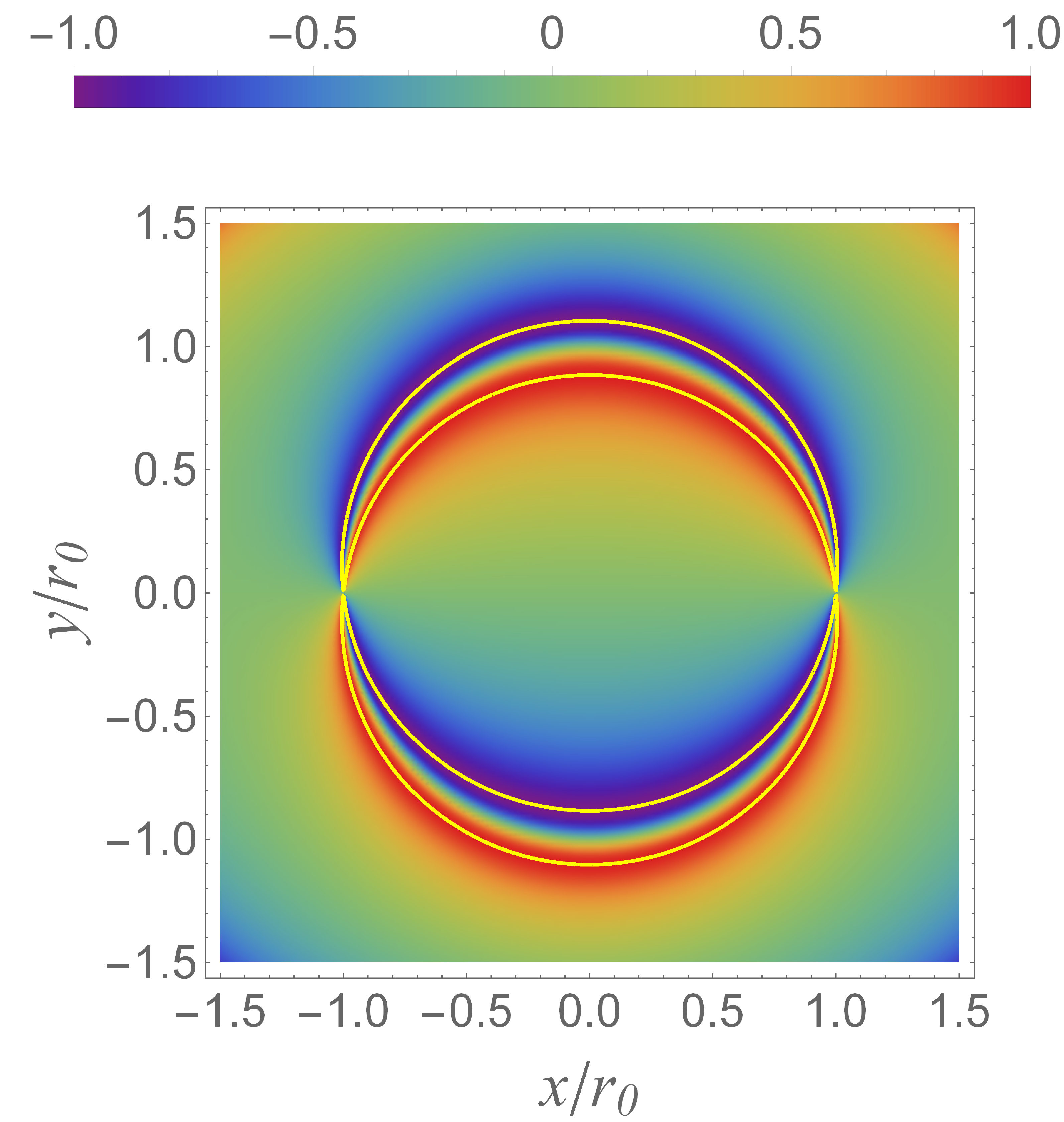}}
 \caption{ \label{fig2}
 Circular c lines (yellow curves) calculated from Eq. \eqref{CRBessel}, superimposed to a density plot of the
ratio $S_3/S_0$, evaluated substituting  Eqs. \eqref{Exa1} into Eqs. \eqref{AllStokesPQ}.
The beam parameter are the same as in Fig. \ref{fig1}.}
\end{figure}
Figure \ref{fig1} shows the temporal evolution of $\vE_\text{real}(\br,t)$ as a function of the scaled coordinates $z/z_0$ and $y/\rho_0$ at $x=0$, where  $\rho_0 = a/{k_r}$ and $a \approx 2.405$ denotes the first zero of  $J_0$.
Substituting the  real and imaginary parts of $\psi(\br)$  into Eq. \eqref{CR1}, we find the implicit equation  of the c lines for the field \eqref{ERealBessel}:
\begin{align}\label{CRBessel}
\cot \vartheta_0 = \frac{y}{\sqrt{x^2+y^2}} \frac{J_1  \bigl({k_r} \sqrt{x^2+y^2} \bigr)}{J_0  \bigl({k_r} \sqrt{x^2+y^2} \bigr)} .
\end{align}
This equation can be numerically inverted to find the curves (c lines)  along which \eqref{CRBessel} becomes an identity. These are illustrated in Fig. \ref{fig2}.

\subsection{Evanescent waves}\label{Evanescentfield}

Consider now the evanescent field  of an inhomogeneous plane wave exponentially decaying in the positive $y$-direction and propagating along the $z$-axis \cite{Jackson}:
\begin{align}\label{Evanescent}
\psi(\br) = \exp\left(- k_t y  \right) \exp \left( i z k_z  \right) ,
\end{align}
where $ k_t = k  \sinh \vartheta$, $ k_z = k  \cosh \vartheta$ and the real parameter $\vartheta$ fixes the scale of inhomogeneity. For $\vartheta = 0$ the wave is purely propagating. Exactly as in the previous case,
the electric field vector is found by substituting
\begin{equation}
\eta =  \frac{k_t}{k} \exp\left(-  k_t y  \right), \qquad
\zeta =  \frac{k_z}{k} \exp\left(-  k_t y  \right) ,
\end{equation}
into Eq. \eqref{vApw3}.
Transverse spin AM density and local wave vector are given, as usual, by  Eq. \eqref{SE2} and Eq. \eqref{KX}, respectively. For the observable electric vector field we obtain from Eq. \eqref{RealE}
\begin{align}\label{ERealBessel3}
\vE_\text{real}(\br,t) =  & \;  \exp\left(-  k_t y  \right)  \Bigl[ \hat{\by} \frac{k_z}{k} \cos \left( z k_z - \omega t \right)  \Bigr. \nonumber \\
& \; \Bigl. + \hat{\bz} \, \frac{k_t}{k}  \sin \left( z k_z - \omega t \right) \Bigr] .
\end{align}
This equation clearly shows the photonic wheel features of evanescent waves \cite{BliokhEvanescent}.
For the sake of completeness, we have calculated the Stokes parameters in the meridional $yz$-plane. The result is
\begin{equation}\label{StokEvan}
\begin{split}
S_0  =  & \; \exp \left( - 2 k y \sinh \vartheta \right)\cosh( 2 \vartheta) , \\
S_1  =  & \; -  \exp \left( - 2 k y \sinh \vartheta \right) ,   \\
S_2  =  & \; 0, \\
S_3  =  & \;  \exp \left( - 2 k y \sinh \vartheta \right)\sinh( 2 \vartheta) ,
\end{split}
\end{equation}
\\
where $S_3/S_0 =  \tanh(2 \vartheta)$. This means that for $\vartheta \gtrsim 1$ the electric field has (almost) transverse circular polarization  uniformly over the $yz$-plane.
%
%\vfill

%
\section{Concluding remarks}\label{Conclusions}

In conclusion, we have presented a perfectly general theory of light waves carrying an electric field  rotating around an axis perpendicular to the propagation direction, thus producing nonzero transverse spin AM density.
 {To define longitudinal and transverse components of the spin angular density, we have assumed that a  propagation direction can be unambiguously identified either locally or globally. In the first case, the direction of light at any point $\mathbf{r}$ is identified by the local wave vector $\mathbf{k}_E(\mathbf{r})$. In the second case, the propagation direction is fixed by the mean wave vector $\langle \mathbf{k}_E(\mathbf{r}) \rangle$ defined as the integral of the local wave vector over the cross-section of the beam.}
The novelty of our approach resides in its general character, which provides for a conceptual unifying view of seemingly different polarization wave phenomena.
The success of such unification is made manifest in the several examples reported in this paper. Last but not least, our treatment reveals a somewhat hidden and intriguing connection between light with circular polarization and the theory of complex functions.

\section*{Acknowledgments}

PB acknowledges financial support by the Alexander von Humboldt Foundation (Feodor Lynen fellowship) and by the Canada Excellence Research Chair (CERC) in Quantum Nonlinear Optics.

%\bibliography{biblio}

%

\end{document}